\documentclass[conference]{IEEEtran}
\IEEEoverridecommandlockouts
\usepackage{cite}
\usepackage{amsmath,amssymb,amsfonts}
\usepackage{algorithmic}
\usepackage{graphicx}
\usepackage{textcomp}
\usepackage{xcolor}
\def\BibTeX{{\rm B\kern-.05em{\sc i\kern-.025em b}\kern-.08em
    T\kern-.1667em\lower.7ex\hbox{E}\kern-.125emX}}
\usepackage{graphicx}
\usepackage{tikz}
\usetikzlibrary{shapes.geometric, arrows}
\tikzstyle{startstop} = [rectangle, rounded corners, minimum width=3cm, minimum height=1cm,text centered, draw=black, fill=red!30]
\tikzstyle{io} = [trapezium, trapezium left angle=70, trapezium right angle=110, minimum width=3cm, minimum height=1cm, text centered, draw=black, fill=blue!30]
\tikzstyle{process} = [rectangle, minimum width=3cm, minimum height=1cm, text centered, draw=black, fill=orange!30]
\tikzstyle{decision} = [diamond, minimum width=3cm, minimum height=1cm, text centered, draw=black, fill=green!30]
\tikzstyle{arrow} = [thick,->,>=stealth]
\usetikzlibrary{calc}
\usepackage{relsize}
\usepackage{pgfplots}
\usetikzlibrary{plotmarks}
\usetikzlibrary{arrows}
\usetikzlibrary{spy,backgrounds}
\usepackage{tikz-3dplot}
\usetikzlibrary{positioning,chains,fit,shapes,calc}
\usetikzlibrary{pgfplots.groupplots}
\tikzset{fontscale/.style = {font=\relsize{#1}}}
\usepackage{amsthm}

\usepackage[ruled]{algorithm2e}
\pgfplotsset{table/search path={data}}
\pgfplotsset{compat=1.3}
\usepackage{ragged2e}
\newcommand\gauss[2]{1/(#2*sqrt(2*pi))*exp(-((x-#1)^2)/(2*#2^2))} 
\pgfplotsset{width=7cm,compat=1.14}
\usepackage{todonotes}

\begin{document}

\title{A Reconstruction-Computation-Quantization (RCQ) Approach to Node Operations in LDPC Decoding}

\author{	
		\makebox[.5\linewidth]{Linfang Wang, Richard D. Wesel}\\\textit{University of California, Los Angeles}\\	\textit{Department of Electrical and Computer Engineering}\\					\{lfwang,wesel\}@ucla.edu%
		\and
		\makebox[.5\linewidth]{Maximilian Stark, Gerhard Bauch}\\ \textit{Hamburg University of Technology} \\	\textit{Institute of Communications}\\\{maximilian.stark,bauch\}@tuhh.de\\
}
\maketitle

\begin{abstract}
In this paper, we propose a finite-precision decoding method that features the three steps of Reconstruction, Computation, and Quantization (RCQ).
Unlike Mutual-Information-Maximization Quantized Belief Propagation (MIM-QBP), RCQ can approximate either belief propagation or Min-Sum decoding.
One problem faced by MIM-QBP decoder is that it cannot work well when the fraction of degree-2 variable nodes is large. However, sometimes a large fraction of degree-2 variable nodes is necessary for a fast encoding structure, as seen in the IEEE 802.11 standard and the DVB-S2 standard. 
In contrast, the proposed RCQ decoder may be applied to any off-the-shelf LDPC code, including those with a large fraction of degree-2 variable nodes.
Our simulations show that a $4$-bit  Min-Sum RCQ decoder delivers frame error rate (FER) performance around $0.1dB$ of full-precision belief propagation (BP)  for the IEEE 802.11 standard LDPC code in the low SNR region.
The RCQ decoder actually outperforms full-precision BP in the high SNR region because it overcomes elementary trapping sets that create an error floor under BP decoding.  
This paper also introduces Hierarchical Dynamic Quantization (HDQ) to design the non-uniform quantizers required by RCQ decoders.  HDQ is a low-complexity design technique that is slightly sub-optimal.  Simulation results comparing HDQ and an optimal quantizer on the symmetric binary-input memoryless additive white Gaussian noise channel show a loss in mutual information between these two quantizers of less than $10^{-6}$ bits, which is negligible for practical applications.
\end{abstract}


\section{Introduction}
{\let\thefootnote\relax\footnote{{This research is supported by National Science Foundation (NSF) grant CCF-1911166 Physical Optics Corporation (POC) and SA Photonics. Any opinions, findings, and conclusions or recommendations expressed in this material are those of the author(s) and do not necessarily reflect views of the NSF, POC, or SA.}}}

Low-Density Parity-Check (LDPC) codes have been widely used in wireless communication and NAND flash system because of its excellent error correction capability. Typically, the massage passing algorithms, which are used to decode LDPC codes, involve accurate number representation. In order to make LDPC code practical, quantization is inevitable. However, uniformly quantizing messages with too low precision will deteriorate the decoder's performance greatly. 

Recently, non-uniform quantization LDPC decoders have raised researchers' interests because of their excellent performance with low precision and coarse quantization. One way to realize non-uniform quantization LDPC decoders is to design lookup tables (LUT) for variable nodes and/or check nodes. In \cite{Planjery2012-FAID}, a Finite Alphabet Iterative Decoder (FAID) is proposed to overcome the error floor of LDPC code under binary symmetric channel (BSC). 
On the other hand, aiming to minimizing the performance degradation in the water fall region, \cite{Romero2016-IBkurk} proposed a Mutual-Information-Maximization LUT (MIM-LUT) decoder. The
MIM-LUT decomposes the actual node operation into  a series of cascaded binary-input-single-output LUTs at the variable and the check node. 
In \cite{Lewandowsky2018-IB}, Lewandowsky et al. proposed the Information-Optimum decoder, which is also called Information Bottleneck (IB) decoder. 
Stark et al. extended the ideas from \cite{Romero2016-IBkurk} and \cite{Lewandowsky2018-IB} and developed message alignment (MA) in \cite{Stark2018-IBMA,Stark2019-IBPBRL} such that IB decoders work also on irregular LDPC codes with arbitrary degree distribution. 
In  \cite{Meidlinger2015-IBMinRe,Meidlinger2017-IBMinIrr},  the Min-LUT decoders were proposed, which replace the LUTs in the check node by a discrete, cluster-based Min-Sum operation. 
The Min-LUT decoder cannot perform well if the fraction of degree-2 variable nodes is large, thus suitable LDPC codes for Min-LUT decoders need careful optimization \cite{Meidlinger2017-IBMinIrr}.

The other way to realize non-uniform quantization is designing quantization parameters that maximizes mutual information between the source and quantized messages. In \cite{Lee2005-RFQThorpe}, Jason Kwok-San Lee and Jeremy Thorpe proposed a non-uniform BP decoder, which is implemented based only on simple mappings and fixed-point additions. 
Unfortunately, the authors did not provide a systematic way to find those mapping parameters.
Recently, He et al. in \cite{He2019-RFQCai} provided a systematic way to find mappings by implementing density evolution and dynamic programming quantization \cite{Kurkoski2014-QuanKur}, and propose MIM-QBP. They also extended MIM-QBP to the irregular LDPC code. However, similar to Min-LUT, MIM-QBP
also faces the problem that it does not work well when the fraction of degree-2 variable nodes  in the LDPC code is large \cite{He2019-RFQCai}.

Even though both Min-QBP and MIM-LUT can have an excellent decoding performance by optimizing edge distribution to lower the fraction of degree 2 variable node, sometimes it is necessary to consider LDPC code with large part of degree 2 variable node. For an example, in the IEEE 802.11 standard the rate $1/2$ LDPC code, half variables nodes has degree 2 for the purpose of fast encoding \cite{mankar2016reduced80211}. 

In this work, we generalize the structure in \cite{Lee2005-RFQThorpe} and propose a finite-precision decoding method that features the three steps of Reconstruction, Computation, and Quantization (RCQ). Unlike MIM-QBP and Min-LUT, RCQ can be applied on any off-the-shelf LDPC codes, including those with larger fraction of degree-2 variable nodes, such as IEEE 802.11 code. The main contributions in this paper include:
\begin{itemize}
    \item We proposed generalized RCQ decoder structure. Unlike the work in \cite{Lee2005-RFQThorpe,He2019-RFQCai}, RCQ decoder can be an approximation of either BP decoder (\textit{bp-RCQ}) or Min-Sum decoder (\textit{ms-RCQ}).  
    \item We designed an efficient sub-optimal quantization scheme, which is called Hierarchical Dynamic Quantization (HDQ), for symmetric binary-input discrete memorelyess channel (BIDMC). HDQ is used for channel quantization and RCQ decoder construction. 
    \item We used HDQ to implement Mutual Information Maximization Discrete Density Evolution (MIM-DDE), and showed that the RCQ decoder is a result of MIM-DDE.
    \item We designed a 4 bit \textit{bp-RCQ} decoder for IEEE 802.11 standard rate $1/2$ LDPC code for theoretical interests. Simulation shows that a $4$-bit \textit{bp-RCQ} decoder delivers frame error rate (FER) performance less than $0.1dB$ of full-precision BP.
    \item We designed a 4 bits \textit{ms-RCQ} decoder for IEEE 802.11 standard rate $1/2$ LDPC code for practical implementation interests. Simulations show that a $4$-bit \textit{ms-RCQ} decoder delivers frame error rate (FER) performance around $0.1dB$ of full-precision belief propagation (BP) in the low SNR region. 
    The RCQ decoder actually outperforms full-precision BP in the high SNR region because it overcomes elementary trapping sets that create an error floor under BP decoding.  
\end{itemize}

The remainder of this paper is organized as follows: In Sec. \ref{sec: RFQ}, we give the description and notations for the RCQ decoder. A hierarchical dynamic quantization algorithm is proposed in Sec. \ref{sec: HDQ}. Mutual information maximization Discrete Density Evolution is introduced in Sec. \ref{sec: MIM-DDE}. This section also describes how to design RCQ decoders given an LDPC ensemble. Simulation results and discussion are given in Sec. \ref{sec: DIS}. Finally, Sec. \ref{sec: CON} concludes our work.


\section{Reconstruction Computation Quantization Decoding Structure}\label{sec: RFQ}
Message passing algorithms update messages between variable nodes and check nodes in an iterative manner either until a valid codeword is found, or a predefined maximum number of iterations, $I_T$, is achieved. The updating procedure contains two steps: 1) computation of the output , 2) message exhange of the output between neighboring nodes. We call the messages with respect to the computation \textit{internal message}, and the messages passed over the edges of the Tanner graph \textit{external message}. In \cite{Lee2005-RFQThorpe}, the authors proposed a LDPC decoder structure where the internal message has a higher precision than external message.
In this work, we generalize their structure and propose a decoding framework that features three steps of Reconstruction, Computation and Quantization.

\begin{figure}
	\centering
	    \begin{tikzpicture}[scale=0.5]
    \draw[black, very thick] (3,1) rectangle (5,2)  node [pos=.5]  {\footnotesize $R(\cdot)$};
    \draw[black, very thick] (3,2.5) rectangle (5,3.5)  node [pos=.5]  {\footnotesize $R(\cdot)$};
    \draw[black, very thick] (3,4) rectangle (5,5)  node [pos=.5]  {\footnotesize $R(\cdot)$};
    \draw[black, very thick] (3,5.5) rectangle (5,6.5)  node [pos=.5]  {\footnotesize $R(\cdot)$};
    \draw [->] (2,1.5) -- (3,1.5);
    \draw [->] (2,3) -- (3,3);
    \draw [->] (2,4.5) -- (3,4.5);
    \draw [->] (2,6) -- (3,6);
    \node [left] at (2,1.5) {$u_{n}$};
    \node [left] at (2,3) {$u_{3}$};
    \node [left] at (2,4.5) {$u_{2}$};
    \node [left] at (2,6) {$u_{1}$};
    \draw [->] (5,1.5) -- (7,1.5);
    \draw [->] (5,3) -- (7,3);
    \draw [->] (5,4.5) -- (7,4.5);
    \draw [->] (5,6) -- (7,6);
    \node [above] at (6,1.5) {$r_{n}$};
    \node [above] at (6,3) {$r_{3}$};
    \node [above] at (6,4.5) {$r_{2}$};
    \node [above] at (6,6) {$r_{1}$};
    \draw[black, very thick] (7,1) rectangle (11,6.5)   node [pos=.5]  { $\mathcal{F}(\cdot)$};
    \draw [->] (11,4) -- (13,4);
    \node [above] at (12,4) {$r_{out}$};
    \draw[black, very thick] (13,3.5) rectangle (15,4.5)  node [pos=.5]  {\footnotesize  $Q(\cdot)$};
    \draw [->] (15,4) -- (16,4);
    \node [right] at (16,4) {$u_{out}$};
    \end{tikzpicture}
	\caption{RCQ Decoding Structure Illustration}
    \label{fig: RFQ_str}
\end{figure}
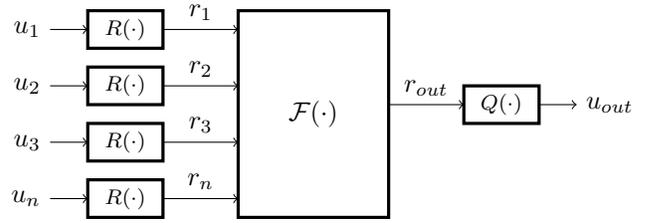

As illustrated in Fig. \ref{fig: RFQ_str}, RCQ decoder consists the following three parts:
\subsubsection{Reconstruction}
Reconstruction $R(\cdot):\mathbb{F}_2^m\rightarrow\mathbb{F}_2^n (m<n)$  maps external message $u_i$ to internal messages $r_i$. We denote channel reconstruction by $R^{ch}$, denote variable node reconstruction and check node reconstruction at iteration $i$ by $R^{c}_i$ and $R^{v}_i$, respectively.
\subsubsection{Computaion}
$\mathcal{F}(\cdot): \mathbb{F}_2^n\rightarrow\mathbb{F}_2^n$ is used to calculate outcoming message. 
We denote the variable node function and check node function by $\mathcal{F}^v$ and $\mathcal{F}^c$, respectively. 
$\mathcal{F}^v$ sums up all incoming messages. 
$\mathcal{F}^c$ has different implementation, we denote check node operation in BP (i.e. hyperbolic-tangent operation) and Min-Sum (we use stantadrd Min-Sum in our work) decoder by $\mathcal{F}^c_{bp}$ and $\mathcal{F}^c_{ms}$. 

\subsubsection{Quantization}
A quantizer $Q:  \mathbb{F}_2^n\rightarrow\mathbb{F}_2^m$ quantizes $n$ bits internal message to $m$ bit external message. A $m$ bits Quantizer $Q$ is determined by $2^m-1$ thresholds $\mathbf{th}=\{th_1,...,th_{2^m-1}\}$ and 
\begin{align}
   Q(i)=\left\{\begin{matrix}
 0&i\leq th_1 \\ 
 2^m-1&i> th_{2^m-1}\\
 j & th_{j}< i\leq th_{j+1} 
\end{matrix}\right.
\label{equ: 10}
\end{align}

We denote channel quantization by $Q^{ch}$, denote check node quantization and variable node quantization at $i^{th}$ iteration by $Q^{c}_i$ and $Q^v_i$ respectively.

RCQ decoder precision can be fully described by a three tuple $\left(m,n^c,n^v\right)$, which represents external message precision, check node internal message precision and variable node internal message precision.
We use notation $\infty$ to denote floating point representation.
\section{Hierarchical Dynamic Quantization}\label{sec: HDQ}
Like most non-uniform quantization LDPC decoders, designing RCQ decoder involves quantization that maximizes mutual information. 
Kurkoski in \cite{Kurkoski2014-QuanKur} proposed a dynamic programming method to find optimal quantizer for BIDMC with complexity $\mathcal{O}(M^3)$, where $M$ is cardinality of channel output.
Dynamic programming quantization is proved to be optimal, however quantization becomes impractical when $M$ is large. 
To mitigate computation complexity, different low-complexity near-optimal algorithms are proposed.
In \cite{Tal2011-QuanVardy}, Tal developed an annealing quantization algorithm with complexity $\mathcal{O}(M\log(M))$ for quantizing symmetric BIDMC . 
In \cite{Lewandowsky2018-IB} Lewandowsky J. improved sequential Information Bottleneck algorithm (sIB) to quantize symmetric BIDMC . The computation of IB algorithm is $\mathcal{O}(tM)$, where $t$ is the number of trials. As a machine learning algorithm, IB algorithm requires multiple trials for a guaranteed a satisfying result.  
In this work, we propose an efficient $m$ bit quantization algorithm for symmetric BIDMC with complexity $\mathcal{O}(mM)$.

Consider code bits $x\in\{0,1\}$ in a binary LDPC codeword are modulated by Binary Phase Shift Keying (BPSK), i.e. $s(x)=-2x+1$, and transmitted by Additive Gaussian White Noise (AWGN) channel. Assume $x$ obeys uniform distribution and noise variance is $\sigma^2$, the joint probability density function between $x$ and received signal $y$, $p(x,y|\sigma)$ is
\begin{align}
    p(x,y|\sigma)&=\frac{1}{2\sqrt{2\pi\sigma^2}}e^{\frac{\left(y-s(x)\right)^2}{2\sigma^2}}.
\end{align}
Since HDQ is designed under BIDMC, we first uniformly quantize $p(x,y|\sigma)$ into $M$ levels and denote the joint probability mass function (p.m.f.) by $P(X,Y),X=\{0,1\}, Y=\{0,...,M-1\}$. We denote $P(X=i,Y=i)$ by $P(X_i,Y_j)$ for simplicity.

 A $m$ bit Quantizer $Q^{ch}$ aims to maximizing mutual information between $X$ and quantized value $T$ \cite{Kurkoski2014-QuanKur} :
\begin{align}
    \arg\max_{Q\in\mathcal{Q}} I(X;T).
\end{align}

\begin{figure}
	\centering
	\begin{tikzpicture}[scale = 0.8]
\begin{axis}[every axis plot post/.append style={
  mark=none,domain=-4:4,samples=50,smooth},
clip=false,
axis y line=none,
axis x line*=bottom,
ymin=0,
xtick=\empty,
]
\draw[->, thick] (-5,0)--(5,0) node[right]{$x$};
\draw[->, thick] (0,0)--(0,0.5) node[above]{$p(y|x)$};
\draw[very thick, dashed, color = red] (0,0)--(0,0.45);
\node [below, color = red] at (0,0) {{$a_2$}};
\draw[very thick, dashed, color = blue] (0.9400,0)--(0.9400,0.45);
\node [below, color = blue] at (0.9400,0) {{$a_3$}};
\draw[very thick, dashed, color = blue] (-0.9400,0)--(-0.9400,0.45);
\node [below, color = blue] at (-0.9400,0) {{$a_1$}};
\draw[very thick, dashed, color = gray] (4,0)--(4,0.45);
\node [below, color = gray] at (4,0) {{$a_4$}};
\draw[very thick, dashed, color = gray] (-4,0)--(-4,0.45);
\node [below, color = gray] at (-4,0) {{$a_0$}};
\addplot [color=black] {\gauss{1}{0.8944}};
\addplot [color=black] {\gauss{-1}{0.8944}};
\draw[ thick, dashed, color = red ] (2.5,0.55)--(3.2,0.55) node[right]{bit level 0};
\draw[ thick, dashed, color = blue ] (2.5,0.51)--(3.2,0.51) node[right]{bit level 1};
\node [above, color = red] at (-2.5,0.44) {0};
\node [above, color = blue] at (-2.15,0.44) {0};
\node [above, color = red] at (-0.6,0.44) {0};
\node [above, color = blue] at (-0.30,0.44) {1};

\node [above, color = blue] at (2.5,0.44) {1};
\node [above, color = red] at (2.25,0.44) {1};
\node [above, color = blue] at (0.6,0.44) {0};
\node [above, color = red] at (0.30,0.44) {1};

\end{axis}
\end{tikzpicture}
	\caption{HDQ method illustration: Quantizing symmetric BI-AWGNC observation into 2 bit messages}
	\label{fig: HDQ}
\end{figure}
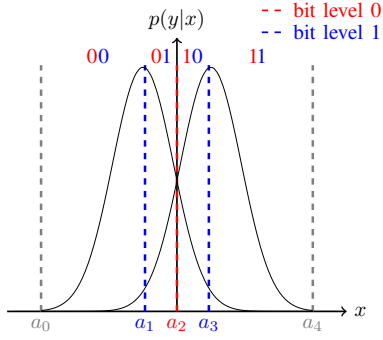
\begin{figure}
	\centering
	    
\begin{tikzpicture}[scale = 1.8]

\node[circle,draw=blue, fill=white, inner sep=0pt,minimum size=5pt] (b) at (-1,0) {};
\node[circle,draw=blue, fill=white, inner sep=0pt,minimum size=5pt] (b) at (-0.25,0) {};
\node[circle,draw=blue, fill=white, inner sep=0pt,minimum size=5pt] (b) at (-0.5,0) {};
\node[circle,draw=blue, fill=white, inner sep=0pt,minimum size=5pt] (b) at (-0.75,0) {};
\node[circle,draw=black, fill=white, inner sep=0pt,minimum size=5pt] (b) at (0,0) {};
\node[circle,draw=orange, fill=white, inner sep=0pt,minimum size=5pt] (b) at (0.25,0) {};
\node[circle,draw=orange, fill=white, inner sep=0pt,minimum size=5pt] (b) at (0.5,0) {};
\node[circle,draw=orange, fill=white, inner sep=0pt,minimum size=5pt] (b) at (0.75,0) {};
\node[circle,draw=orange, fill=white, inner sep=0pt,minimum size=5pt] (b) at (1,0) {};
\node[circle,draw=orange, fill=white, inner sep=0pt,minimum size=5pt] (b) at (1.25,0) {};
\node[circle,draw=orange, fill=white, inner sep=0pt,minimum size=5pt] (b) at (1.5,0) {};
\node[circle,draw=orange, fill=white, inner sep=0pt,minimum size=5pt] (b) at (1.75,0) {};
\node[below]  at (-0.97,-0.06) {$a_l$};
\node[below]  at (1.81,-0.06) {$a_r$};
\node[below]  at (0.03,-0.06) {$a_i$};
\node[above, color=blue]  at (-0.65,0.06) {\footnotesize{{$P_l$}}};
\node[above, color=orange]  at (1,0.06) {\footnotesize{{$P_r$}}};
\node[above, color=black]  at (0,0.06) {\footnotesize{{$P_m$}}};
\draw[->] (0,0.3)--(0,0.7);
\node[right, color=black]  at (0,0.5) {\footnotesize{\texttt{cost}($P_l,P_m$)$>$\texttt{cost}($P_r,P_m$)}};
\node[above, color=black]  at (0,0.7) {\footnotesize{Stop: Return $a_i$}};
\draw [blue, dashed,rounded corners] (-1.12,-0.08) rectangle (-0.12,0.08);
\draw [orange, dashed,rounded corners] (0.12,-0.08) rectangle (1.87,0.08);
\draw[->] (0,-0.25)--(0,-0.65);
\node[right, color=black]  at (0,-0.45) {\footnotesize{\texttt{cost}($P_l,P_m$)$\leq$\texttt{cost}($P_r,P_m$)}};
\node[circle,draw=blue, fill=white, inner sep=0pt,minimum size=5pt] (b) at (-1,-1) {};
\node[circle,draw=blue, fill=white, inner sep=0pt,minimum size=5pt] (b) at (-0.25,-1) {};
\node[circle,draw=blue, fill=white, inner sep=0pt,minimum size=5pt] (b) at (-0.5,-1) {};
\node[circle,draw=blue, fill=white, inner sep=0pt,minimum size=5pt] (b) at (-0.75,-1) {};
\node[circle,draw=blue, fill=white, inner sep=0pt,minimum size=5pt] (b) at (0,-1) {};
\node[circle,draw=black, fill=white, inner sep=0pt,minimum size=5pt] (b) at (0.25,-1) {};
\node[circle,draw=orange, fill=white, inner sep=0pt,minimum size=5pt] (b) at (0.5,-1) {};
\node[circle,draw=orange, fill=white, inner sep=0pt,minimum size=5pt] (b) at (0.75,-1) {};
\node[circle,draw=orange, fill=white, inner sep=0pt,minimum size=5pt] (b) at (1,-1) {};
\node[circle,draw=orange, fill=white, inner sep=0pt,minimum size=5pt] (b) at (1.25,-1) {};
\node[circle,draw=orange, fill=white, inner sep=0pt,minimum size=5pt] (b) at (1.5,-1) {};
\node[circle,draw=orange, fill=white, inner sep=0pt,minimum size=5pt] (b) at (1.75,-1) {};
\node[below]  at (-0.97,-1.06) {$a_l$};
\node[below]  at (1.81,-1.06) {$a_r$};
\node[below]  at (0.28,-1.06) {$a_{i+1}$};
\node[above, color=blue]  at (-0.50,-0.94) {\footnotesize{{$P_l$}}};
\node[above, color=orange]  at (1.125,-0.94) {\footnotesize{{$P_r$}}};
\node[above, color=black]  at (0.25,-0.94) {\footnotesize{{$P_m$}}};
\draw [blue, dashed,rounded corners] (-1.12,-1.08) rectangle (0.12,-0.92);
\draw [orange, dashed,rounded corners] (0.37,-1.08) rectangle (1.87,-0.92);
\end{tikzpicture}
	\caption{An intermediate step of STS Algorithm}
	\label{fig: STS}
\end{figure}
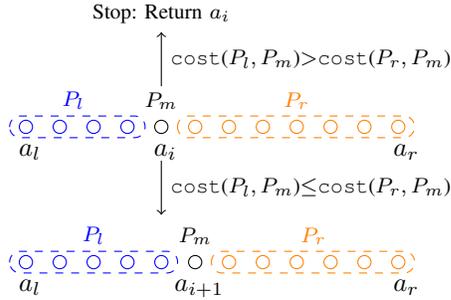

\begin{algorithm}
\label{alg: STS}
\SetKwInOut{Input}{input}\SetKwInOut{Output}{output}\SetKwProg{Init}{initialization:}{}{}
\SetKwProg{Ret}{return}{}{}

\Input{$P(X,Y)$, $a_l$, $a_r$}
\Output{$a_{out}$}

$P_l\leftarrow [P(X_0,Y_{a_l})\quad P(X_1,Y_{a_l})]$\\
$P_m\leftarrow [P(X_0,Y_{a_l+1})\quad P(X_1,Y_{a_l+1})]$\\
$P_r\leftarrow [\sum_{i=a_l+1}^{a_r-1}P(X_0,Y_i) \quad \sum_{i=a_l+1}^{a_r-1}P(X_1,Y_i)]$\\
\For{$i\gets1$ \KwTo $a_r-a_l-2$ }{
    $c_i^l\leftarrow \texttt{cost}(P_l,P_m)$\\
    $c_i^r\leftarrow \texttt{cost}(P_r,P_m)$\\
    \eIf{$c_i^l<c_i^r$}
    {
    $P_l\leftarrow P_l+P_m$\\
    $P_r\leftarrow P_r-P_m$\\
    $P_m\leftarrow [P(X_0,Y_{a_l+i+1})\quad P(X_1,Y_{a_l+i+1})]$
    }
    {
        \KwRet $a_l+i+1$
    }
    }

\KwRet $a_r-1$
\caption{Sequential Thresholds Searching (\textit{STS}) }
\end{algorithm}

\begin{algorithm}
\label{alg: HDQ}
\SetKwInOut{Input}{input}\SetKwInOut{Output}{output}\SetKwProg{Init}{initialization:}{}{}

\Input{$\Pr\left(X,Y\right), X\in\{0,1\}, Y\in\{0,...,N-1\}$; $m$}
\Output{ $P(X,T)$, $Q$, $R$}
$a_0\leftarrow 0$\\
$a_{N}\leftarrow N-1$\\
\For{$i\gets0$ \KwTo $m-1$ }{
    \For{$j\gets0$ \KwTo $2^{i-1}-1$ }{
    $a_{\frac{T}{2^i}(j+\frac{T}{2})}\leftarrow \texttt{STS}\left(a_{\frac{T}{2^{i}}j},a_{\frac{T}{2^{i}}\left(j+1\right)}\right)$
    }
    }
$P(X_i,T_j)\leftarrow \sum_{k=0}^{a_j-1}P(X_i,T_k)$\\
$th_i\leftarrow\log\frac{P(X_0,Y_{a_i})}{P(X_1,Y_{a_i})}$ \\
$R(i) = \log\frac{P(X_0,T_i)}{P(X_1,T_i)}$
\caption{Hierarchical Dynamic Quantization }
\end{algorithm}

Lemma 1 and Lemma 2 in \cite{Romero2016-IBkurk} simplifies finding an optimal $m$ bit quantizer to finding $2^m-1$ boundaries $\{a_1,..., a_{2^m-1}\}$. Even so, jointly optimizing $2^m-1$ boundaries still has a large searching space. Hence, instead of optimizing thresholds jointly, HDQ algorithm determines these boundaries \textit{bit level by bit level}. Figure. \ref{fig: HDQ} illustrates how HDQ quantizes symmetric BI-AWGNC output into 2 bit levels :
\begin{itemize}
    \item initialize:  $a_0$ and $a_4$.
    \item bit level 0: determine $a_2$, $a_0<a_2<a_{4}-1$,
    \item bit level 1 : fix $a_2$ and determine $a_1$ and $a_3$, $a_0<a_1<a_2-1$ and $a_2<a_3<a_4-1$.
\end{itemize}

Note that $a_1$ and $a_3$ are independently optimized, it is easy to show that the solution of $a_1$ is independent to the solution of $a_3$. A similar idea is also used in optimizing progressive reads for flash memory cells\cite{Wong2019-eh}.We borrow the metric of Information Bottleneck algorithm and develop a sequential threshold searching algorithm (STS) to find $a_i$. Given $a_l$ and $a_r$, $r>l$ and  starting from $a_{l+1}$, STS sequentially calculates the merging costs that $a_i$ is merged into left or right cluster until left merging cost is larger than right merging cost. Fig. \ref{fig: STS} shows an intermediate step of STS. Merging cost is defined as mutual information loss when merging two probabilities together(Ref \cite{Lewandowsky2018-IB}, Eq(10)) . Full description of STS and HDQ algorithm are given in Algorithm \ref{alg: STS} and \ref{alg: HDQ}, respectively. 


\begin{figure}
    \centering
    \includegraphics[scale=0.43
    ]{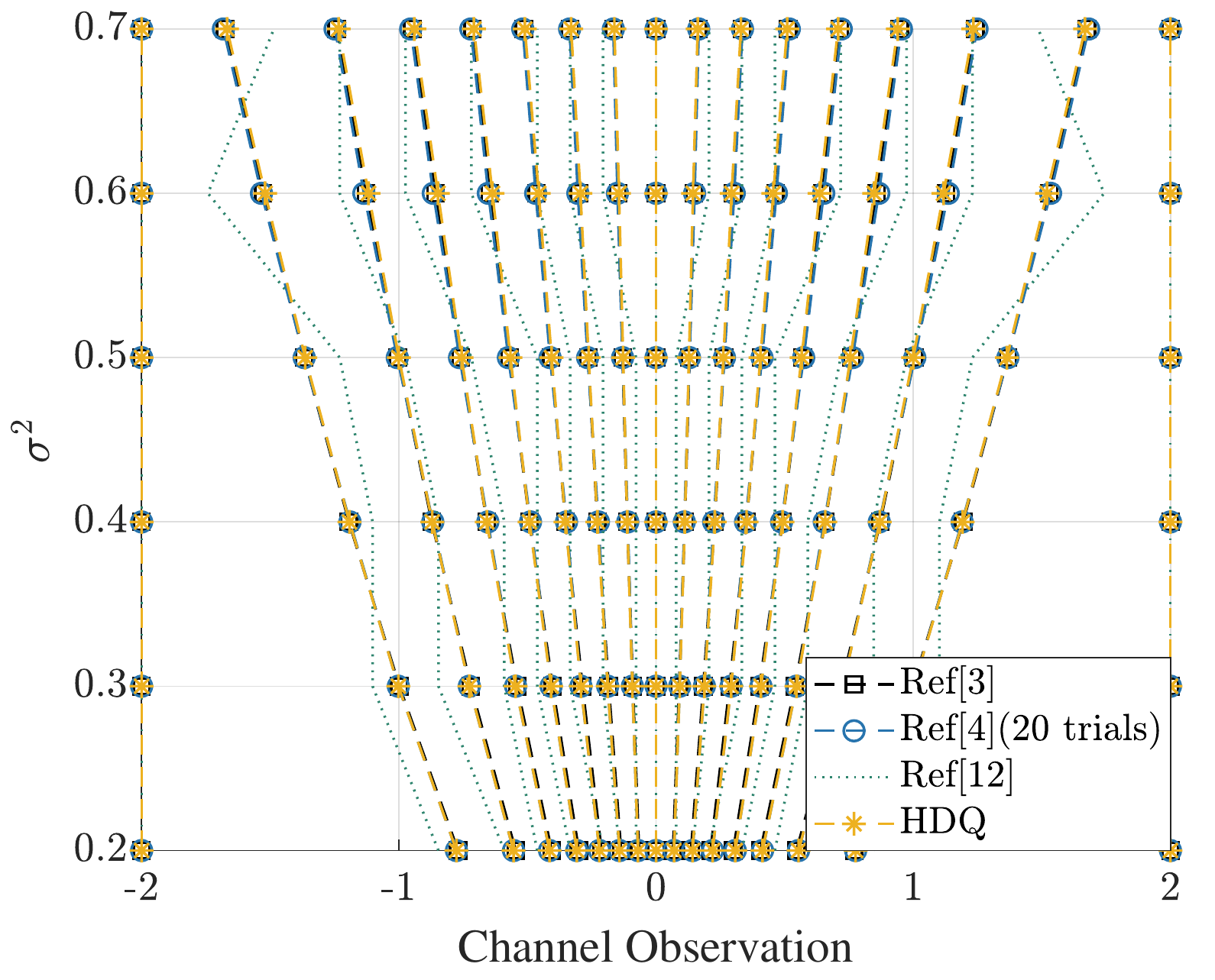}
    \caption{  quantization regions for channel  output of BI-AWGNC under different $\sigma^2$}
    \label{fig: quan_result}
\end{figure}

Fig. \ref{fig: quan_result} shows 4 bits quantization regions for channel output of BI-AWGNC under different $\sigma^2$. We examined four different quantization algorithms. 
Simulation shows that improved sIB algorithm and HDQ algorithm has a quantization result very close to the optimal dynamic programming algorithm. 
Annealing quantization algorithm deviates from the optimal solution to different extent under different $\sigma^2$. 
We use $I^{dp}(X;T)$ to denote the mutual information between $X$ and quantized value $T$, obtained by optimal dynamic programming quantizer and use $I^{sub}(X;T)$ to represent mutual information obtained through sub-optimal quantizers. Therefore, we can quantitatively evaluate the performance of each sub-optimal algorithm by:
\begin{align}
    \Delta I_{sub} = I^{dp}(X;T)-I^{sub}(X;T).
\end{align}

\begin{figure}
    \centering
    \includegraphics[scale=0.38
    ]{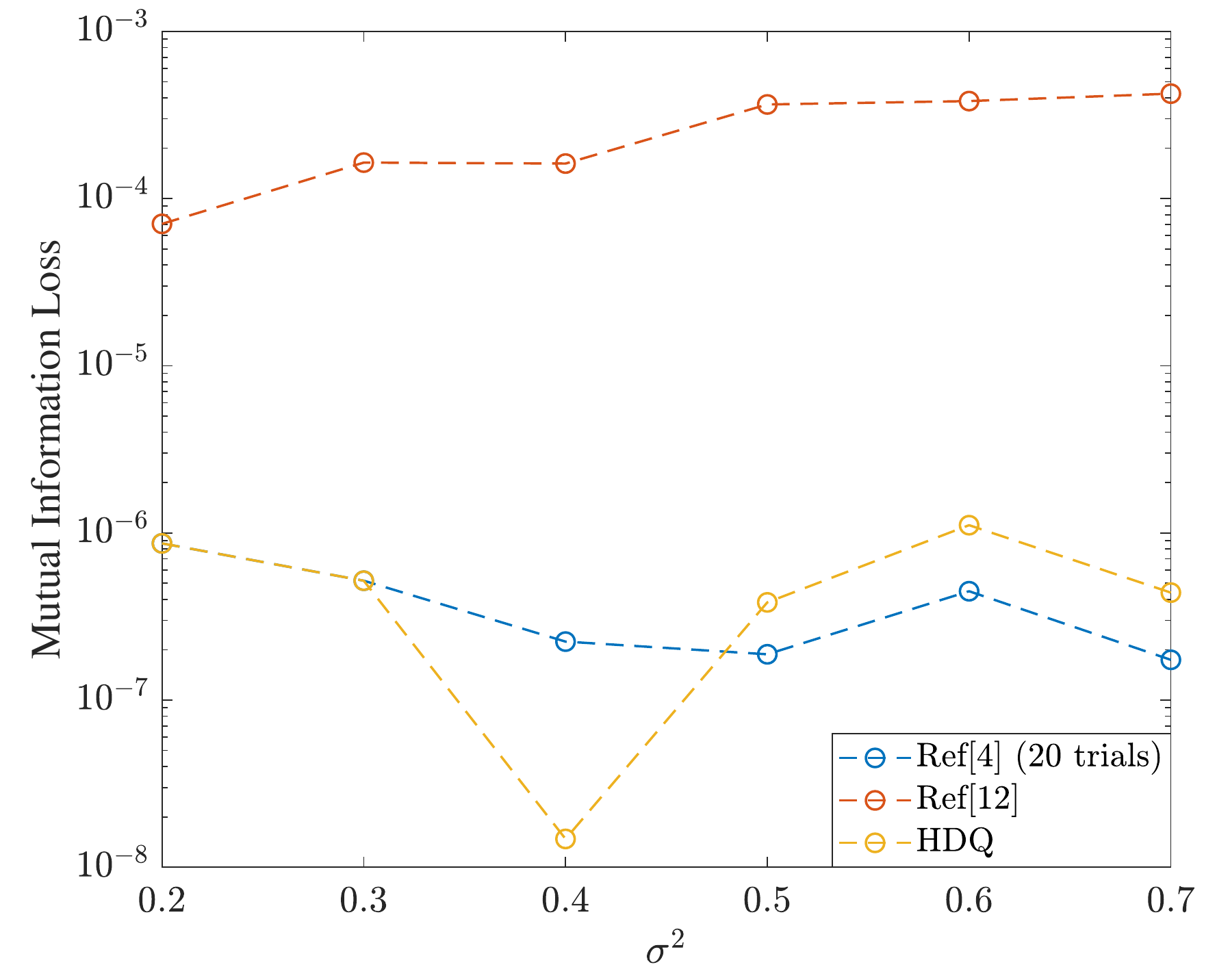}
    \caption{Difference of mutual information loss between each sub-optimal quantizer and optimal quantizer}
    \label{fig: mi_loss}
\end{figure}

Fig. \ref{fig: mi_loss} gives $\Delta I_{sub}$ of each sub-optimal quantizer. Simulation shows that all three sub-optimal quantizer yields very similar mutual information with optimal quantizer. However, we can still see that compared with annealing quantization, sIB algorithm and HDQ has a quantization result more close to optimal quantizer because the $\Delta I_{sub}$ is around $10^{-6}$ for both sIB and HDQ. 

In the next section, we will use HDQ to conduct mutual-information-maximization discrete density evolution and construct RCQ decoder.

\section{Mutual Information Maximization\\ Discrete Density Evolution}\label{sec: MIM-DDE}
RCQ decoder is a result of quantized density evolution : By quantizing the joint p.m.f. between code bits and message from variable node or check node, $R^c_i$,$R^v_i$,$Q^c_i$,$Q^v_i$ can be constructed correspondingly. To differ our discrete density evolution with the one using uniform quantization\cite{Sae-Young_Chung2001-DDE0045}, we call our density evolution Mutual-Information-Maximization Discrete Density Evolution (MIM-DDE).

\subsection{MIM-DDE at check node}
 Denote the joint p.m.f between incoming message $T$ and code bit $X$ from $i^{th}$ variable node by $P^{v,i}(X,T)$, $X=\{0,1\}$, $T=\{0,...,2^m-1\}$. Based on the independence assumption in the density evolution \cite{Richardson2001-de}, we have:
\begin{align}
    P^{v,i} (X,T)=P^v(X,T),\quad i=0,...,d_c-1 
\end{align}
where $d_c$ is check node degree. At check node, the code bit corresponding to output is the XOR sum of code bits corresponding to all inputs. By denoting:

{\footnotesize
\begin{align}
     P^{v,a}(X,T)\circledast P^{v,b}(X,T)\triangleq\sum_{\substack{m,n:\\m\bigoplus n=k} }P^{v,a}(X_m,T)P^{v,b}(X_n,T),
\end{align}
}where $m,n,k\in\{0,1\}$, the joint p.m.f between code bit corresponded to output and input messages, $P^c_{out}(X,\mathbf{T})$, can be represented by:
\begin{align}
    P^c_{out}(X,\mathbf{T})&=P^{v,0}(X,T)\circledast ...\circledast P^{v,d_c-2}(X,T)\\
    &=P^{v}(X,T)\circledast ...\circledast P^{v}(X,T)\\
    &\triangleq P^v(X,T)^{\circledast(d_c-1)}, \label{equ: check_opt_prob}
\end{align}
where $\mathbf{T}$ is a vector containing all incoming $d_c-1$ messages.
Eq.(\ref{equ: check_opt_prob}) gives p.m.f. update when $\mathcal{F}^c_{bp}$ is implemented at the check node.

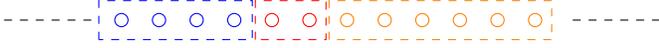
\begin{figure}
	\centering
	\begin{tikzpicture}[scale = 2.0]

\node[circle,draw=blue, fill=white, inner sep=0pt,minimum size=5pt] (b) at (-1,0) {};
\node[circle,draw=blue, fill=white, inner sep=0pt,minimum size=5pt] (b) at (-0.25,0) {};
\node[circle,draw=blue, fill=white, inner sep=0pt,minimum size=5pt] (b) at (-0.5,0) {};
\node[circle,draw=blue, fill=white, inner sep=0pt,minimum size=5pt] (b) at (-0.75,0) {};
\node[circle,draw=red, fill=white, inner sep=0pt,minimum size=5pt] (b) at (0,0) {};
\node[circle,draw=red, fill=white, inner sep=0pt,minimum size=5pt] (b) at (0.25,0) {};
\node[circle,draw=orange, fill=white, inner sep=0pt,minimum size=5pt] (b) at (0.5,0) {};
\node[circle,draw=orange, fill=white, inner sep=0pt,minimum size=5pt] (b) at (0.75,0) {};
\node[circle,draw=orange, fill=white, inner sep=0pt,minimum size=5pt] (b) at (1,0) {};
\node[circle,draw=orange, fill=white, inner sep=0pt,minimum size=5pt] (b) at (1.25,0) {};
\node[circle,draw=orange, fill=white, inner sep=0pt,minimum size=5pt] (b) at (1.5,0) {};
\node[circle,draw=orange, fill=white, inner sep=0pt,minimum size=5pt] (b) at (1.75,0) {};
\draw[black, dashed] (2.0,0)--(2.6,0);
\draw[black, dashed] (-1.2,0)--(-1.8,0);
\draw [blue, dashed] (-1.15,-0.13) rectangle (-0.13,0.13);
\draw [red, dashed] (-0.11,-0.13) rectangle (0.36,0.13);
\draw [orange, dashed] (0.38,-0.13) rectangle (1.86,0.13);
\end{tikzpicture}
	\caption{OSA illustration: points are ordered w.r.t. LLR values. Each color represents a cluster and LLR value difference in each cluster is less than $l_s$. }
	\label{fig: OSA}
\end{figure}

In order to keep cardinality of external message same, $P^c_{out}(X,\mathbf{T})$ needs to be quantized to $2^m$ levels. As pointed in \cite{Lewandowsky2018-IB}, $|\mathbf{T}|=2^{m(d_c-1)}$ will be very large when $m$ and $d_c$ is large. For an example, if $d_c=8$ and $m=4$, $|\mathbf{T}|=2.68*10^8$. Hence, directly quantizing $P^c_{out}(X,\mathbf{T})$ is impossible. To mitigate the problem of \textit{cardinality bombing}, we propose an intermediate coarse quantization algorithm called One-Step-Annealing (OSA) quantization without sacrificing mutual information. Note that Eq. (\ref{equ: check_opt_prob}) can be calculate in a recursive way and each step takes two input:
\begin{align}
\label{equ: check_recur}
    P_{out}^{c}(X,\mathbf{T})^{\circledast i}=P^v(X,T)^{\circledast (i-1)}\circledast P^v(X,T) 
\end{align}
We observe that, in each step, output of Eq.(\ref{equ: check_recur}) have some entries with very close log likelihood ration (LLR) value. By merging entries whose LLR difference is small enough, mutual information loss is negligible.  Hence, OSA simply merges entries whose LLR values difference is less than a threshold $l_s$, and the output of OSA will be the input of next p.m.f calculation step, i.e.:
\begin{align}
\label{equ: check_osa}
    P^{v}(X,T)^{\circledast i}= \texttt{OSA}(P^v(X,T)^{\circledast (i-1)},l_s)\circledast P^v(X,T) .
\end{align}
We take $l_s\in[10^{-4},10^{-3}]$ in our simulation. Fig. \ref{fig: OSA} shows an illustration of OSA and full description of OSA algorithm is given in Algorithm.\ref{alg: OSA}. The following table shows $|\mathbf{T}|$ after we implement OSA and choose different $l_s$. The example we showed has the parameter $m=4$, $d_c=8$. The result shows that OSA greatly decreases the output cardinality, and based on our simulation, mutual information losses under these three $l_s$ are all less than $10^{-7}$.
\begin{table}[h]
\centering
\begin{tabular}{|c|c|c|c|c|}
\hline
$l_s$                                                         & $0$           & $10^{-4}$    & $5*10^{-4}$ & $10^{-3}$  \\ \hline
\begin{tabular}[c]{@{}c@{}}$|\mathbf{T}|$\end{tabular} & $2.68*10^{8}$ & $3.3*10^4$ & $1.7*10^3$  & $1.3*10^3$ \\ \hline
\end{tabular}
\end{table}

For a regular LDPC code with check node degree $d_c$, HDQ is implemented to quantize $\mathbf{T}$ into a $m$ bit message. We denote joint p.m.f. between code bit $x$ and quantized value $T$ by $P^c(X,T)$. As a result of HDQ, $Q^c$ and $R^v$ in this iteration are constructed.

Unlike regular LDPC code, irregular LDPC code has different node types, we denote the check node edge distribution by $\rho(x)=\sum_{i=2}^{d_{c,max}}\rho_ix^{i-1}$. To update $P^c(X,T)$ and construct $Q^c_k$ and $R_k^v$ for irregular LDPC code, we need to quantize:
\begin{align}
    P^c_{out}(X,\mathbf{T})&=\sum_{i=2}^{d_c}\rho_i P^c(X,T)^{\circledast(i-1)}
\end{align}

Due to space limitation, we refer \cite{Meidlinger2015-IBMinRe} to Min-Sum operation. Note that Min-Sum operation doesn't change the cardinality of output, this implies for \textit{ms-RCQ}: 
\begin{enumerate}
    \item $m=n^c$.
    \item $R^c$ is not required. We can map $2^m$ messages to $(-2^m-1,...,-1,1,...,2^m-1)$ and then implement $\mathcal{F}^c_{ms}$. We can also implement a single LUT to realize the min-sum operation.

\end{enumerate}

\begin{algorithm}
\label{alg: OSA}
\SetKwInOut{Input}{input}\SetKwInOut{Output}{output}\SetKwProg{Init}{initialization:}{}{}

\Input{$\Pr\left(X,Y\right), X\in\{0,1\}, Y\in\{0,...,N-1\}$; $l$}
\Output{$\Pr(X,T)$}
$j\leftarrow0$\\
$\Pr(X_0,T_j)\leftarrow P(X_0,Y_0)$\\
$\Pr(X_1,T_j)\leftarrow P(X_1,Y_0)$\\
$l_s\leftarrow  \log\frac{\Pr(X_0,Y_0)}{\Pr(X_1,Y_0)}$

\For{$i\gets1$ \KwTo $N-1$ }{
      \eIf{$(\log\frac{P(X_0,T_i)}{P(X_1,T_i)}-l_s)\leq l$}
    {
    $P(X_0,T_j)\leftarrow\Pr(X_0,T_j)+\Pr(X_0,Y_i)$\\
    $P(X_1,T_j)\leftarrow\Pr(X_1,T_j)+\Pr(X_1,Y_i)$
    }
    {
    $j\leftarrow j+1$\\
    $\Pr(X_0,T_j)\leftarrow\Pr(X_0,Y_i)$\\
    $\Pr(X_1,T_j)\leftarrow\Pr(X_1,Y_i)$\\
    $l_s\leftarrow\log\frac{\Pr(X_0,Y_i)}{\Pr(X_1,Y_i)}$
    }
    }
\caption{One Step Annealing Algorithm (OSA) }\label{sec: MIMDDE}
\end{algorithm}
\subsection{MIM-DDE at variable node}

Variable node sums the LLR messages from channel observation and neighboring check nodes. By denoting: 
\begin{align}
P^{c,a}(X,T)\boxdot P^{c,b}(X,T)&=\frac{1}{P(X)}P^{c,a}(X,T)P^{c,b}(X,T),
\end{align}
the joint p.m.f between code bit $X$ and incoming message combination $\mathbf{T}$, $P^v_{out}(X,\mathbf{T})$, given variable node degree $d_v$,  can be expressed by:
\begin{align}
    P^v_{out}(X,\mathbf{T})=P^{ch}(X,T)\boxdot P^c(X,T)^{\boxdot(d_c-1)},
\end{align}
Similarly, for irregular LDPC code with variable edge degree distribution $\lambda(x)=\sum_{i=2}^{d_{v,max}}x^{i-1}$, $P^v_{out}(X,\mathbf{T})$ is given by:
\begin{align}
     P^v_{out}(X,\mathbf{T})=P^{ch}(X,T)\boxdot \sum_{i=2}^{d_{v,max}}\lambda_i P^c(X,T)^{\boxdot(d_v-1)}.
\end{align}
$P_{out}^v(X,\mathbf{T})$ is then quantized to $2^m$ levels by HDQ. Also, as a result of HDQ, and joint p.m.f between code bit $X$ and quantized messages $T$, $P^v(X,T)$, is updated. $Q^v$ in this iteration and $R^c$ in the next iteration can be built correspondingly. Note that variable node also faces \textit{cardinality bombing} problem, hence $\texttt{OSA}$ is needed in each  recursive step.

Thus, by implementing MIM-DDE, we can iteratively update $P^c(X,T)$, $P^v(X,T)$ and build $Q^c_i$, $Q^v_i$, $R^c_i$ and $R^v_i$, $i=\{0,...,I_T-1\}$. 

In MIM-DDE, we only limit the precision of external messages, i.e. $m$, and keep internal messages, $n^c$ (only for \textit{bp-RCQ}) and $n^v$, full precision. To make internal message precision finite, a uniform $n^c$ (or $n^v$) quantizer is required when implementing $\mathcal{F}^c$(or $\mathcal{F}^v$).

\section{Simulation and Discussion}\label{sec: DIS}

In this section, we build RCQ decoder for IEEE 802.11 standard LDPC code with codeword length $1296$ and rate $0.5$. The edge distribution is:
\begin{align}
    \lambda(x) &= 0.2588x+0.3140x^2+0.0465x^3+0.3837x^{10}, \\
    \rho(x) &= 0.8140x^6 +0.1860 x^7.
\end{align}
The LDPC code we choose has fast encoding structure hence half the variable nodes has degree 2.
The $\frac{E_b}{N_o}$ we used to design RCQ is 0.90 dB for both \textit{bp-RCQ} and \textit{ms-RCQ}. $I_T$ is set to be 50.

\begin{figure}
    \centering
    \includegraphics[scale=0.45
    ]{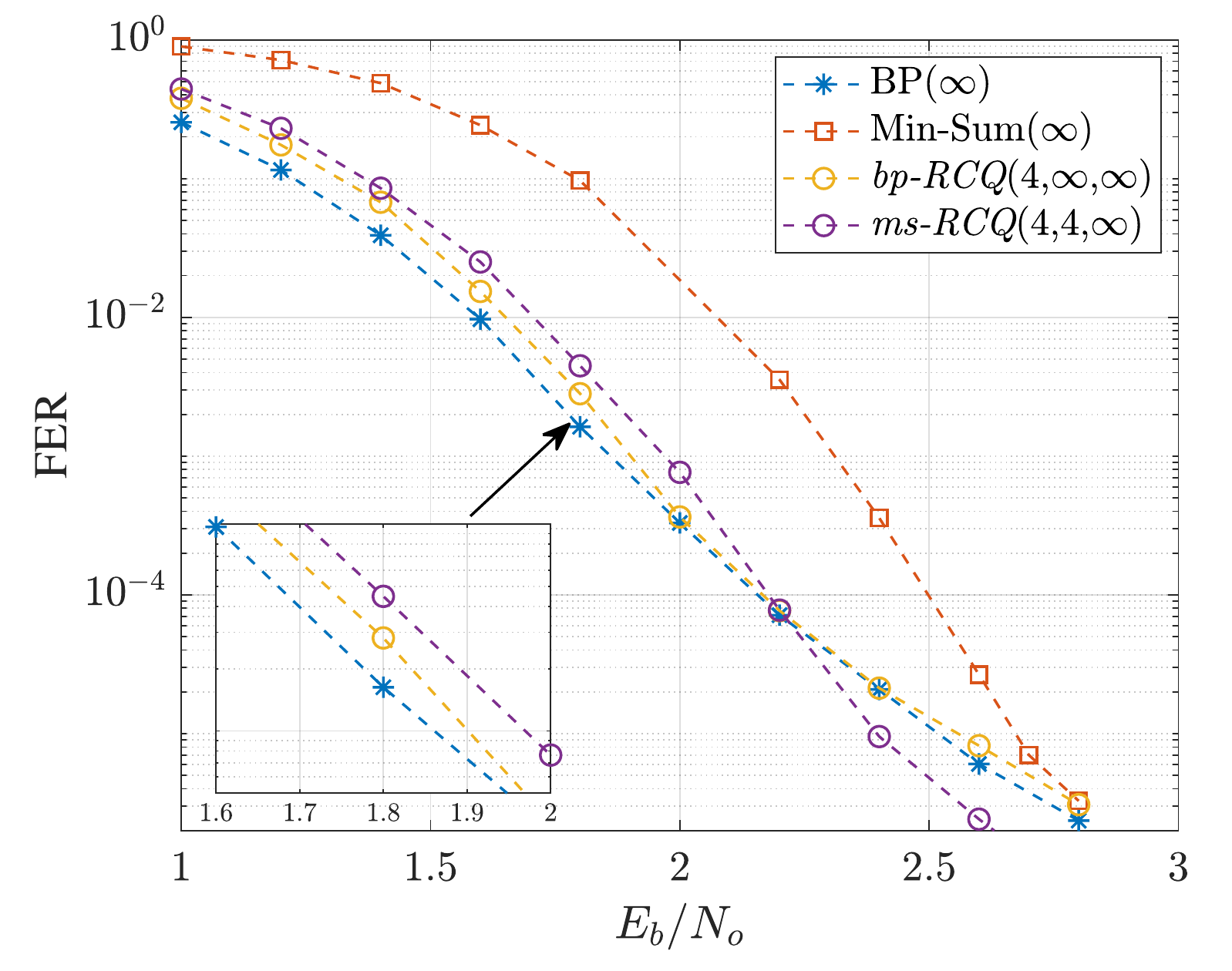}
    \caption{RCQ decoder with full precision internal message}
    \label{fig: RFQ-1}
\end{figure}

\begin{figure}
    \centering
    \includegraphics[scale=0.45
    ]{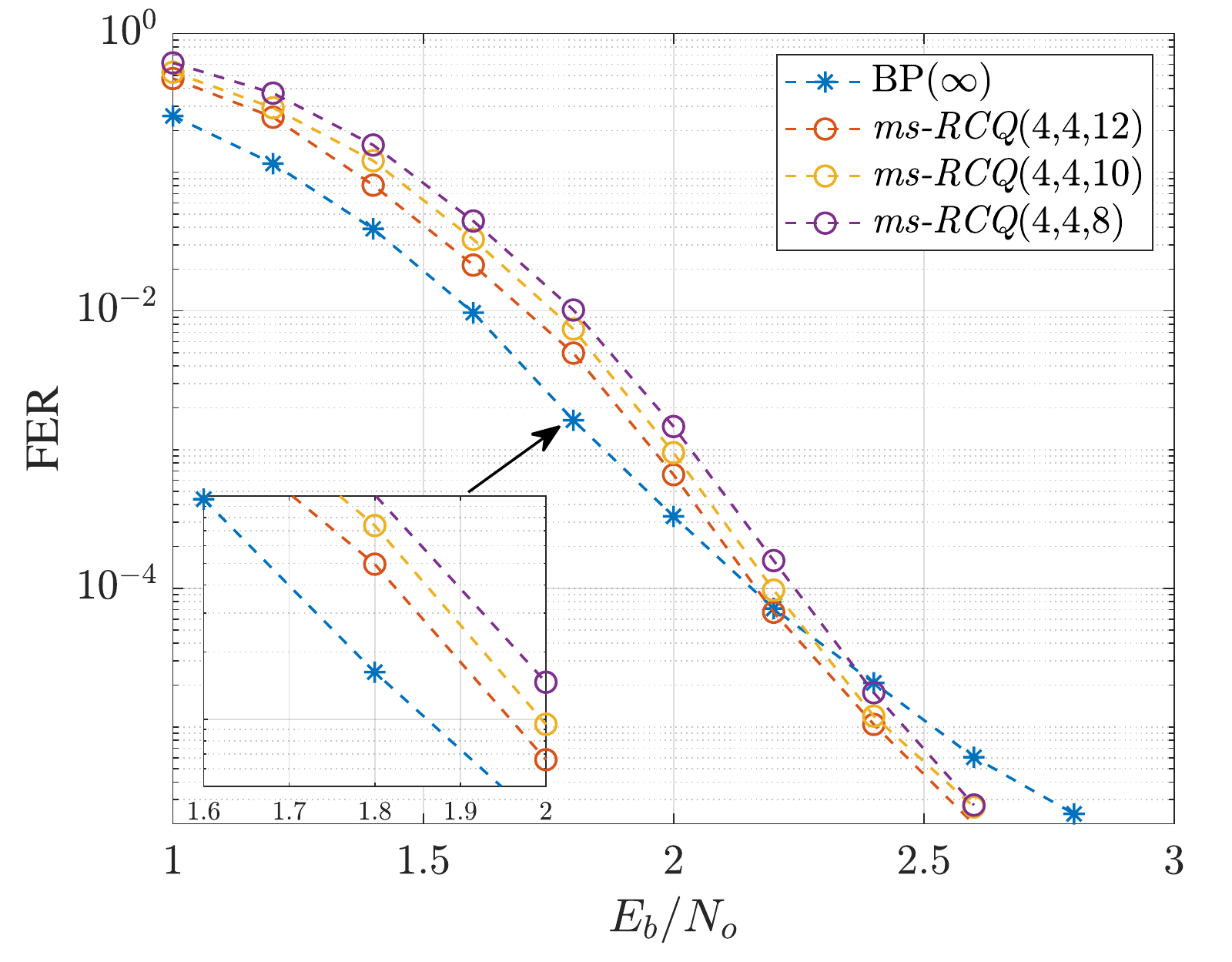}
    \caption{The effect of internal message Quantization for $4$ bits \textit{ms-RCQ}}
    \label{fig: RFQ-2}
\end{figure}
Fig. \ref{fig: RFQ-1} shows the FER simulation result of \textit{bp-RCQ}(4,$\infty$,$\infty$) and \textit{ms-RCQ}(4,4,$\infty$). 
As comparison, we give the performance of BP($\infty$) and Min-Sum ($\infty$). BP decoder performs best, but error floor appears at $2.4$dB. The error floor is due to the existence of trapping sets, which is a result of large degree-2 variable nodes. 
Waterfall of Min-Sum starts from $2.2$ dB, this implies Min-Sum decoder is transparent to trapping set that BP can't overcome.
This phenomena is also observed in \cite{Ryan2009-shulin}. Interestingly, it also reflects on RCQ decoders.
When $\frac{E_b}{N_o}$ is low, compared with BP($\infty$), \textit{bp-RCQ} (4,$\infty$,$\infty$) has a degradation less than 0.1 dB and \textit{ms-RCQ}(4,4,$\infty$) has a degradation around 0.1 dB.
As $\frac{E_b}{N_o}$ increases, \textit{bp-RCQ}(4,$\infty$,$\infty$) behaves similar to BP($\infty$) and appears error floor.
However, \textit{ms-RCQ} (4,4,$\infty$) outperforms BP. We collected noised codewords that BP could not decode under $2.6$dB and fed it into \textit{ms-RCQ}. Simulation result shows \textit{ms-RCQ} can decode $80\%$ of them.

For a purpose of practical use, we are more interested in \textit{ms-RCQ}. Fig.\ref{fig: RFQ-2} gives FER performance of \textit{ms-RCQ} decoder with different $n^v$. When $\frac{E_b}{N_o}<2.2$ dB, \textit{ms-RCQ}(4,4,12) ($5$ bits are assigned to integer part and $7$ bits are assigned to fraction part), \textit{ms-RCQ}(4,4,10) ($5$ bits are assigned to integer part and $5$ bits are assigned to fraction part) and  \textit{ms-RCQ}(4,4,8) ($5$ bits are assigned to integer part and $3$ bits are assigned to fraction part) have a degradation around $0.1$, $0.15$ and $0.2$ dB, compared with BP($\infty$). When $E_b/N_o>2.4$ dB, all three \textit{ms-RCQ} decoders outperforms BP($\infty$).

\section{Conclusion}\label{sec: CON}

In this work, HDQ is proposed to quantize a symmetric binary input discrete channel into $m$ bit levels. Then we use HDQ and MIM-DDE to construct the RCQ decoder. Unlike Mutual-Information-Maximization Quantized Belief Propagation (MIM-QBP), RCQ can approximate either belief propagation or Min-Sum decoding. We use an IEEE 802.11 standard LDPC code to illustrate that the RCQ decoder works well when the fraction of degree 2 variable nodes is large. Simulations show that a $4$-bit \textit{ms-RCQ} decoder delivers frame error rate (FER) performance around $0.1dB$ of full-precision belief propagation (BP) in the low SNR region. The RCQ decoder actually outperforms full-precision BP in the high SNR region because it overcomes elementary trapping sets that create an error floor under BP decoding.

\bibliographystyle{IEEEtran}
\bibliography{conference_101719}

\end{document}